\documentclass[aps,pre,twocolumn,float]{revtex4-1}
\usepackage{amssymb,amsmath,latexsym}
\usepackage{epsfig}
\usepackage[totalwidth=480pt, totalheight=680pt]{geometry}
\def\Fbox#1{\vskip1ex\hbox to 8.5cm{\hfil\fboxsep0.3cm\fbox{%
  \parbox{8.0cm}{#1}}\hfil}\vskip1ex\noindent}  

\begin{document}

\title{Short-time $\beta$-relaxation in glass-forming liquids is cooperative in nature} 

\author{Smarajit Karmakar$^{1}$}
\author{Chandan Dasgupta$^{2,3}$}
\author{Srikanth Sastry$^{3}$}

\affiliation{$^1$ Centre for Interdisciplinary Sciences, Tata Institute 
of Fundamental Research, 21 Brundavan Colony, Narsingi, Hyderabad,
500075, India}
\affiliation{$^2$ Centre for Condensed Matter Theory, Department of Physics,
Indian Institute of Science, Bangalore, 560012, India}
\affiliation{$^3$ Jawaharlal Nehru Centre for Advanced Scientific Research, Bangalore
560064, India.}

%

\begin{abstract}
Temporal relaxation of density fluctuations in supercooled liquids near the glass transition occurs in multiple steps. The short-time $\beta$-relaxation is generally attributed to spatially local processes involving the rattling motion of a particle in the transient cage formed by its neighbors. Using molecular dynamics simulations for three model glass-forming liquids, we show that the $\beta$-relaxation is actually cooperative in nature. Using finite-size scaling analysis, we extract a growing length-scale associated with $\beta$-relaxation from the observed dependence of the $\beta$-relaxation time on the system size. Remarkably, the temperature dependence of this length scale is found to be the same as that of the length scale that describes the spatial heterogeneity of local dynamics in the long-time $\alpha$-relaxation regime. These results show that the conventional interpretation of $\beta$-relaxation as a local process is too simplified and provide a clear connection between short-time dynamics  and long-time structural relaxation in glass-forming liquids.
\end{abstract}
\maketitle

\section{Introduction}
Temporal relaxation of density fluctuations in supercooled liquids near the glass transition occurs in multiple steps. At short times, the temporal autocorrelation of density fluctuations and related correlation functions approach a plateau after a fast initial decay. This part of the relaxation is known as the $\beta$ relaxation. The subsequent long-time decay of the correlation functions from the plateau to zero is known as the $\alpha$ relaxation. The origin of this nonexponential, multi-step relaxation and possible connections between the short-time $\beta$ relaxation and the long-time $\alpha$ relaxation are among the fundamental issues in the study of glassy dynamics.

\begin{figure}[!h]
\begin{center}
\hskip -0.4cm
\includegraphics[scale = 0.350]{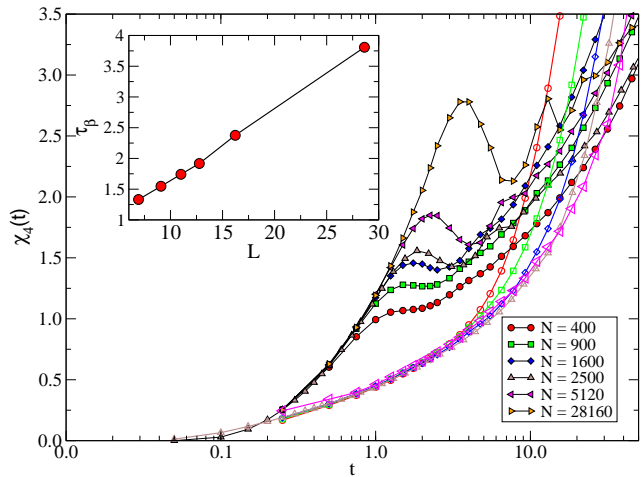}
\caption{Four-point susceptibility $\chi_4(t)$ for different system sizes, plotted
versus time $t$. Note the appearance of a peak at short times. The height of the peak 
increases with increasing system size. The 
filled symbols represent data from molecular dynamics simulations and open symbols
show data from Monte Carlo Simulations. The peak of $\chi_4$ at short
time scales disappears in Monte Carlo simulations. Inset: The height of the
short-time peak of $\chi_4$ is plotted as a function of the length $L$ of the sample.
The height is found to increase linearly with $L$.}
\label{x4BetaPeak}
\end{center}
\end{figure} 
Many approaches to understanding slow relaxation in glass forming
liquids invoke the notion of a growing length scale that governs the
increase of relaxation time scales
\cite{arcmp,adam-gibbs,inhmct,rfotpra,rfotrev,kinetic,kinetic2,Berthier_PRE,Whitelam_Berthier_PRL}.
Experimental and theoretical studies of dynamic heterogeneity in
glass-forming liquids \cite{edigerhet,onuki,harrowell,donati} have lead
to a detailed analysis of length scales that are associated with
spatial correlations of the mobility of particles
\cite{BBBKMR_JCP1_126_2007,BBBKMR_JCP1_126_2007_2,berthier_science,berthier_physreve,zamponi}.
These correlations have been studied through a four-point density
correlation function $g_4 (r,t)$ \cite{chandan,silvio,sharon}, its
Fourier transform - the four-point structure factor $S_4(q,t)$ - and the
associated dynamic susceptibility $\chi_4(t) \equiv \lim_{q \rightarrow 0} S_4(q,t)$ 
\cite {definitions,Berthier_PRE,Whitelam_Berthier_PRL,BBBKMR_JCP1_126_2007,parisik4}. Analytic
predictions for the behaviour of dynamic length scales and
susceptibilities in both short-time ($\beta$) and long-time ($\alpha$)
relaxation regimes have been obtained from inhomogeneous mode coupling
theory (IMCT)~\cite{inhmct,BBBKMR_JCP1_126_2007,BBBKMR_JCP1_126_2007_2}.
These predictions include an initial power-law growth of the
cooperativity length scale in time in the $\beta$ relaxation regime, followed by
a saturation at time scales comparable to the $\alpha$ relaxation time. However, details 
of how the dynamics crosses over from its short-time behaviour to that at
long times and the corresponding crossover in the dynamic length scale(s)
are still not well-understood. There are many ideas and observations that attempt to
relate dynamical features observed at short time scales to long-time
structural relaxation\cite{hallwolynes,buchenau,dyre,harrowellnphys,wyart,starr,leporini,douglas}. Thus,
a proper understanding of relaxation processes in both $\beta$ and
$\alpha$ regimes and their underlying relation (if any) is extremely important
in the overall understanding of glassy dynamics and its rich
phenomenology.

In an earlier study \cite{KDS}, the method of finite-size scaling
(FSS)~\cite{fss}, used extensively for obtaining accurate numerical
results for critical properties near conventional phase transitions,
was used to obtain the length scale $\xi(T)$ associated with dynamical
heterogeneity, by analyzing the size dependence of the dynamic
susceptibility $\chi_4(t)$ and the associated Binder cumulant. It was
subsequently shown that the heterogeneity length obtained in this way is in
good agreement with that obtained by analyzing the four-point
structure factor $S_4(q,t)$ \cite{KDS_PRL10,KDS_comment}. However,
very large system sizes needed to be employed in the analysis of $S_4(q,t)$, and
further difficulties arise from the ensemble dependence of the associated susceptibility
$\chi_4(t)$ \cite{szamel1,szamel2}. Indeed, results \cite{Stein_Andersen_PRL}
claiming to confirm IMCT predictions for $\chi_4(t)$ and the associated length scale
$\xi(T)$ are affected by finite-size effects. Thus, notwithstanding caveats
about its use \cite{szamel1,sarlat}, FSS offers an attractive approach
to study length scales relevant to glassy dynamics. We employ this method in the
present work. 

It was found, however, in \cite{KDS} that the $\alpha$ relaxation
time $\tau_\alpha$ that describes the long-time decay of a two-point
density correlation function does not exhibit the expected dynamical
scaling, in that (a) the time scale decreases instead of increasing with system
size, and (b) a plot of $\tau_\alpha$ scaled by its asymptotic value
for large system sizes {\it vs.} similarly scaled values of
$\chi_4$ does not display a scaling collapse. This was interpreted as
indicating a mixing of activated and non-activated mechanisms of
structural relaxation (see also \cite{bhatta,bbrfot}).  It
has been suggested \cite{bbrfot} that behaviour described by
IMCT, without the influence of activated dynamics, should be expected
instead in the short-time dynamics, in a temperature window above the 
mode coupling transition temperature. 
\begin{figure}[!h]
\begin{center}
\epsfig{file=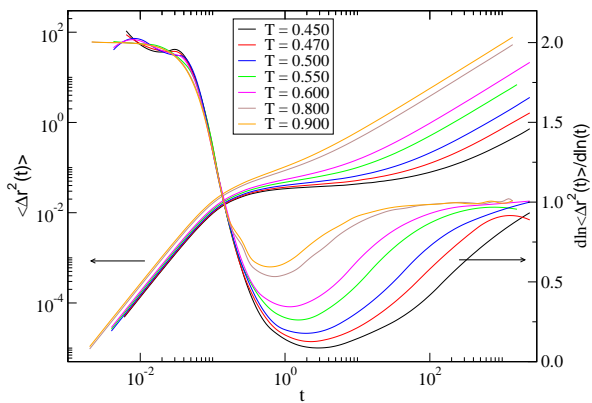,scale=0.250,angle = -90,clip=}
\caption{Mean square displacement (MSD) and the derivative of the
logarithm of MSD with respect to the logarithm of time $t$, shown as a
function of $t$. The minimum of the derivative defines
$\tau_\beta$. (see text for details).}
\label{msd}
\end{center}
\end{figure}

\begin{figure}[!h]
\begin{center}
\epsfig{file=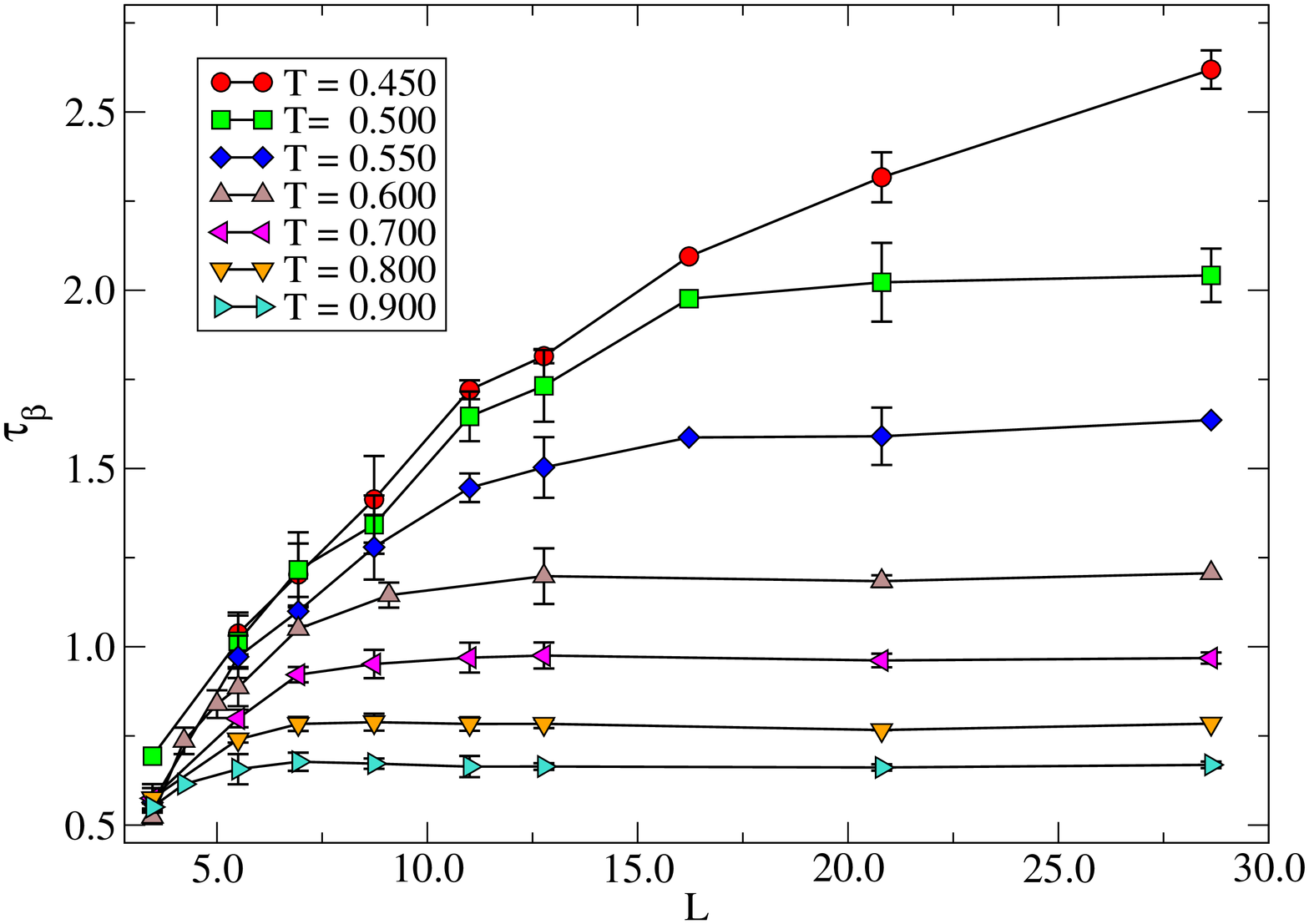,scale=0.30,angle=-90,clip=}
\epsfig{file=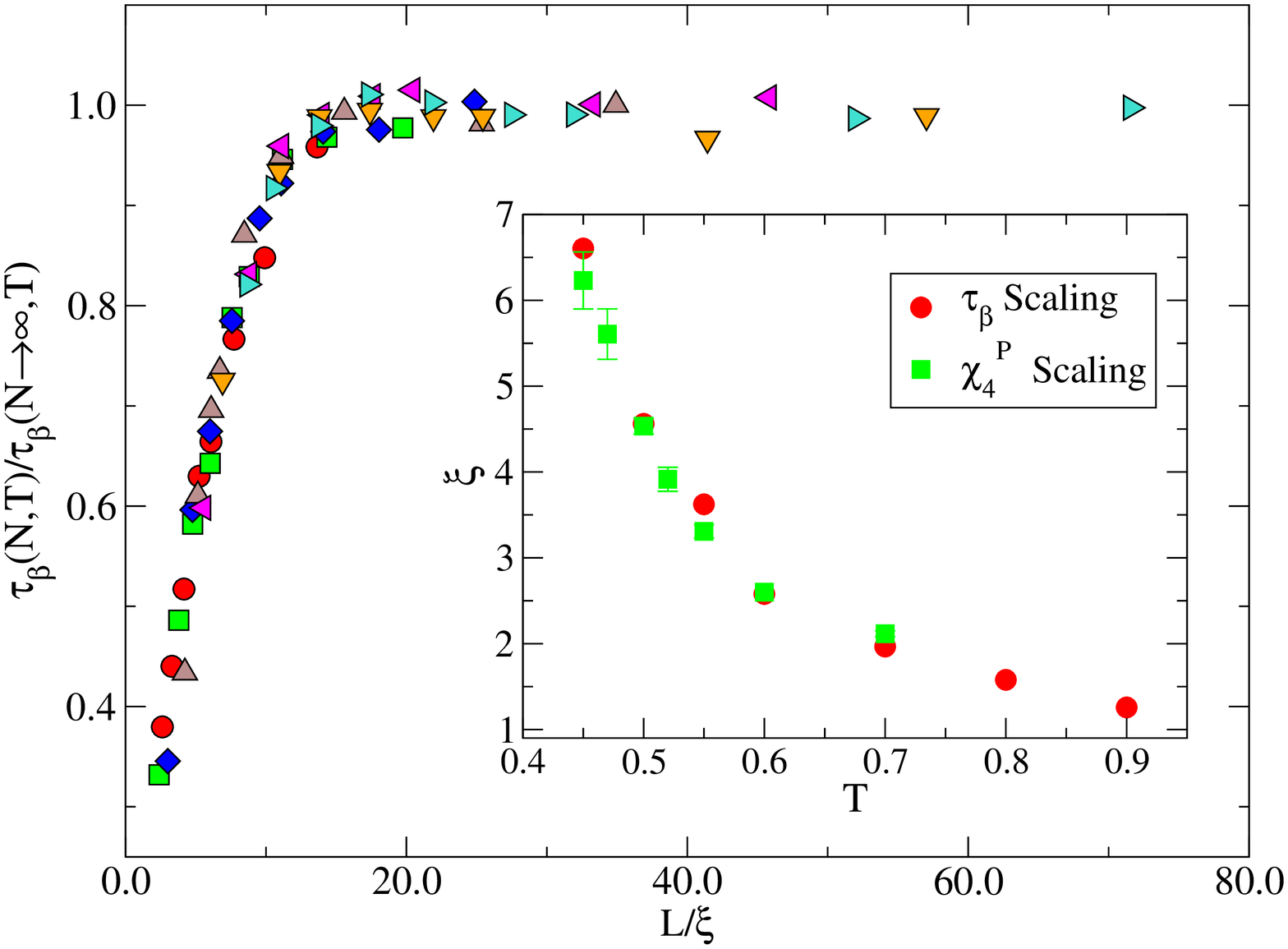,scale=0.330,angle=-90,clip=}
\caption{Top panel: $\tau_\beta$ for the 3dKALJ model, shown as a function of system size for
  different temperatures in the interval $T \in [{0.45, 0.90}]$. $\tau_\beta$ increases with increasing system size, saturating at a value that increases with decreasing temperature. Bottom panel: Scaling collapse of $\tau_\beta$. For each temperature, the
  linear size $L$ is scaled by a length $\xi(T)$ such that data for
  all temperatures collapse to a master curve. Inset: Comparison of the 
  correlation length extracted from the system-size dependence of 
  $\tau_\beta$ with that obtained from the
  $\alpha$ regime \cite{KDS} as a function of temperature. The two
  length scales agree with each other to a good accuracy.}
\label{tauBetaSys}
\end{center}
\end{figure}
\begin{figure*}[!t]
\begin{center}
\epsfig{file=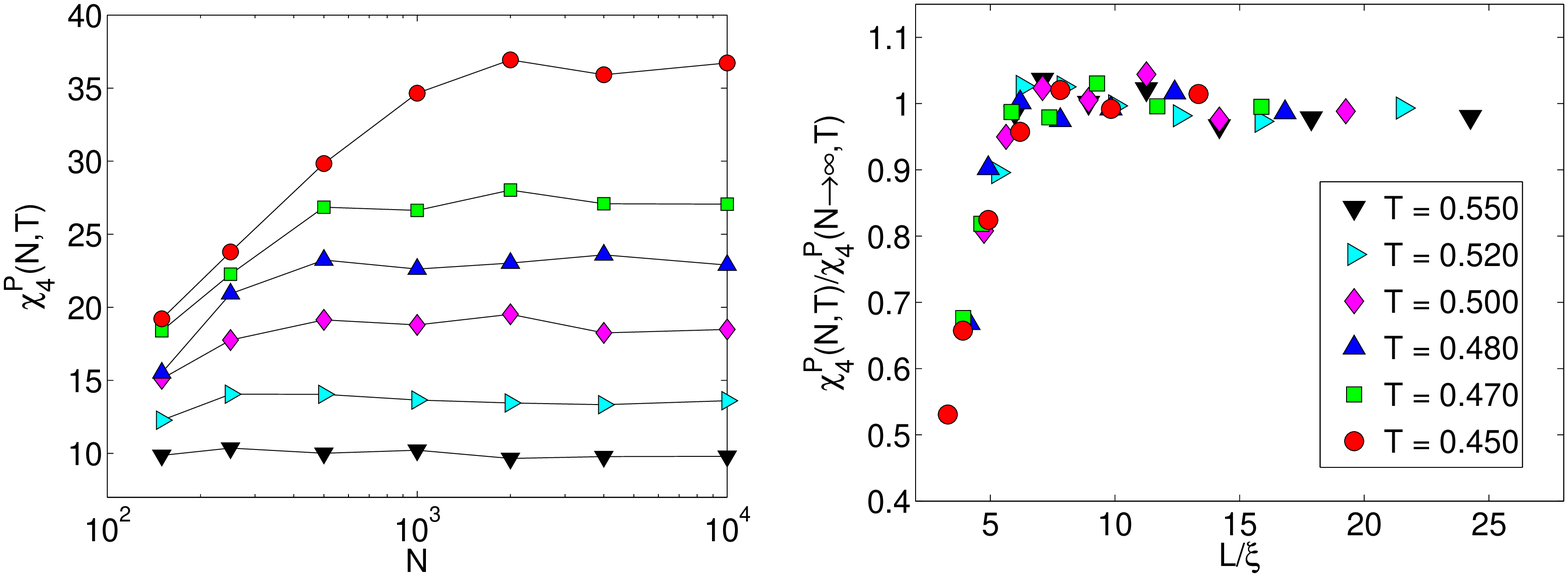,scale=0.50,clip=}
\epsfig{file=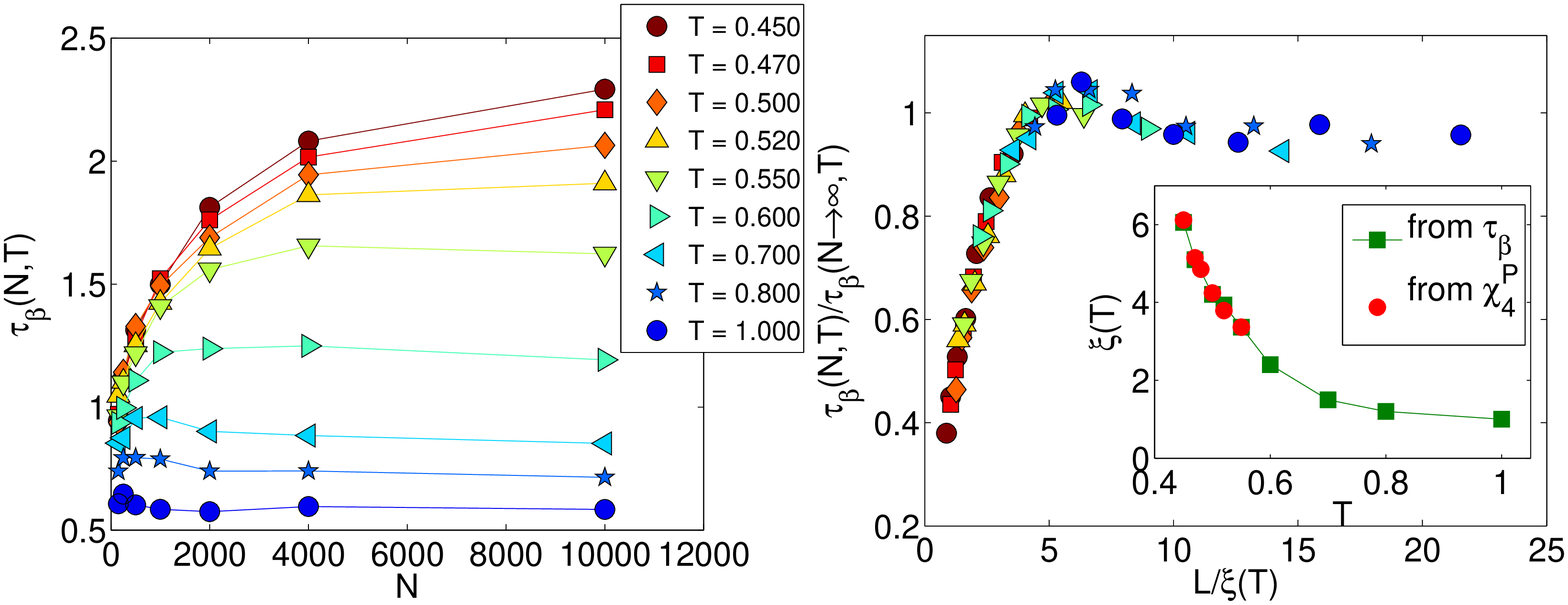,scale=0.530,clip=}
\caption{Top left panel: System-size dependence of $\chi_4^P$, the peak value of $\chi_4(t)$, 
for different temperature for the 3dIPL model system. 
Top right panel: The data collapse of $\chi_4^P(N,T)$ to extract the  dynamic 
heterogeneity length scale. $\chi_4^P(N\to\infty,T)$ is the asymptotic value of 
$\chi_4^P$ in the limit of infinite system size. The corresponding 
length scales are shown in the inset of the bottom right panel of this figure (red 
circles). Bottom left panel: System-size dependence of $\tau_{\beta}$ for all 
the temperature studied, including high temperatures. Bottom right panel: 
Data collapse of $\tau_{\beta}(N,T)$ to obtain the length scale associated 
with the cooperativity in $\beta$-relaxation. Inset: Comparison of the correlation 
length extracted from the system-size dependence of $\tau_\beta$ with that 
obtained from the system-size dependence of $\chi_4^P$, as a function of temperature. 
The two length scales agree with each other to a good accuracy.}
\label{3dIPLresult}
\end{center}
\end{figure*}

In this article, we describe the results of FSS studies
of a short time scale (the $\beta$ relaxation time scale $\tau_\beta$) for three 
generic model glass-formers. The simulation details are provided in Sec. II. 
We compare the estimated length scales with those obtained from FSS 
analysis of $\chi_4^P$, the peak value of $\chi_4(t)$,  which quantifies dynamic heterogeneity 
at the $\alpha$ relaxation time scale, and find that the two length scales 
agree well for all the systems we study. We also find a power-law dependence of the 
$\beta$ time scale on the length scale, in qualitative agreement with the prediction of IMCT. 
These results are described in Sec III. We also perform an analysis of the dependence of 
the behaviour in the $\beta$ regime on the microscopic dynamics, as described in Sec. IV. 
Sec. V contains a discussion of our results and a summary.

\section{Method and Simulation Details}

We study three model liquids in this work. They are:  (i) the three-dimensional Kob-Andersen binary 
Lennard-Jones mixture (referred here as 3dKALJ) \cite{KA}, (ii) a three-dimensional 
system characterized by a repulsive inverse power-law potential (3dIPL) 
considered in Ref. \cite{10PSD} and  (iii) a 
three-dimensional system characterized by a repulsive inverse power-law potential
(3dR10) \cite{12KLP} whose range of interaction is smaller compared 
to that in the first two 
models. For the first two model systems the interaction range covers the second 
neighbouring shell whereas for the third case it is truncated within the first shell. 

\begin{figure*}[!t]
\begin{center}
\epsfig{file=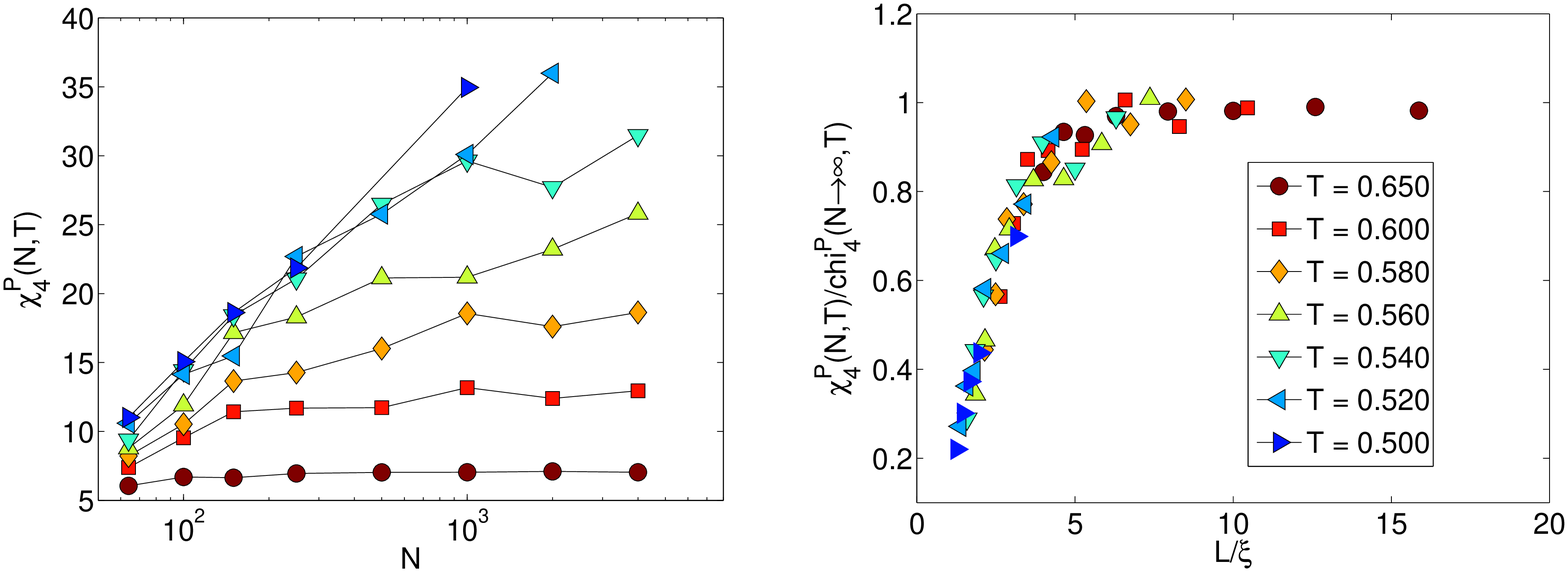,scale=0.44,clip=}
\epsfig{file=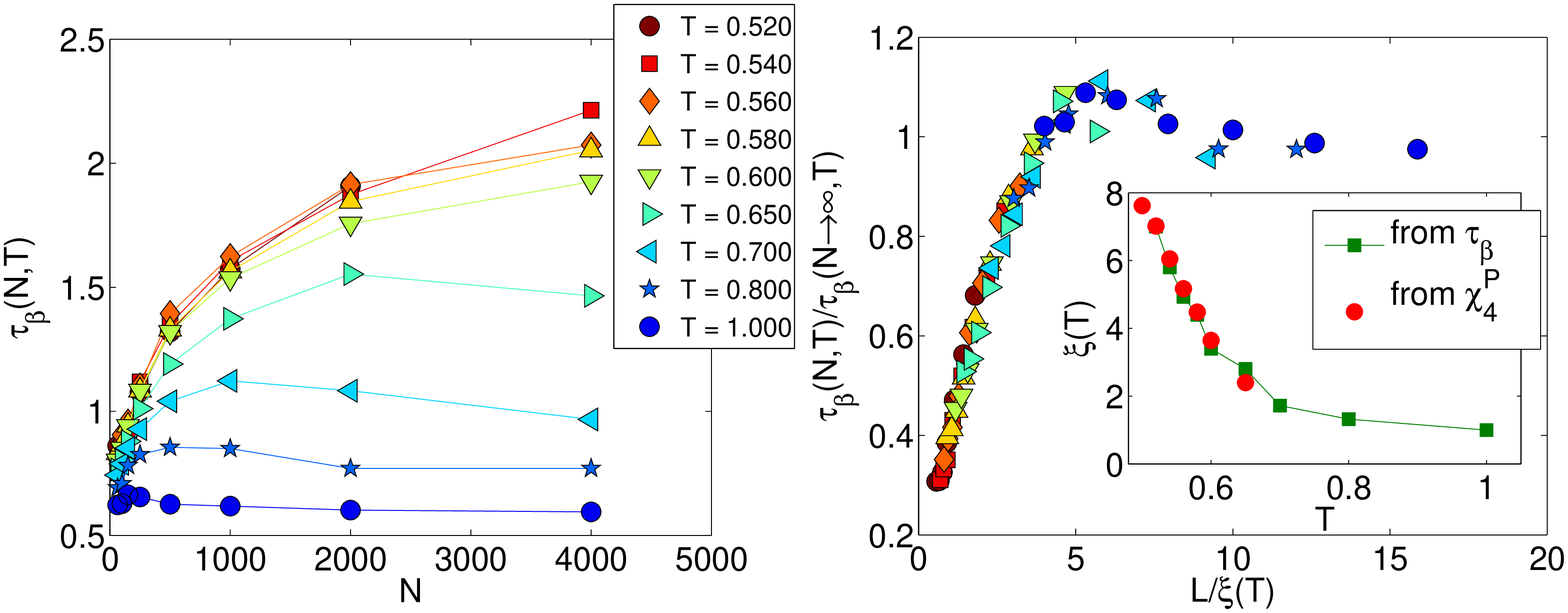,scale=0.5,clip=}
\caption{Top left panel: System-size dependence of $\chi_4^P$, the peak value of $\chi_4(t)$, 
 for different temperature for the 3dR10 model system. 
Top right panel: The data collapse of $\chi_4^P(N,T)$ to extract the  dynamic 
heterogeneity length scale. $\chi_4^P(N\to\infty,T)$ is the asymptotic value of 
$\chi_4^P$ in the limit of infinite system size. The corresponding 
length scales are shown in the inset of the bottom right panel of this figure (red 
circles). Bottom left panel: System-size dependence of $\tau_{\beta}$ for all 
the temperature studied, including high temperatures. Bottom right panel: 
Data collapse of $\tau_{\beta}(N,T)$ to obtain the length scale associated 
with the cooperativity in $\beta$-relaxation. Inset: Comparison of the correlation 
length extracted from the system-size dependence of $\tau_\beta$ with that 
obtained from the system-size dependence of $\chi_4^P$, as a function of temperature. 
The two length scales agree with each other to a good accuracy.}
\label{3dR10result}
\end{center}
\end{figure*}

We have performed a series of simulations, for systems of sizes ranging from
$N = 150$ to $N = 28160$ particles at number density $\rho = 1.20$ of the 
glass-forming Kob-Andersen binary liquid mixture~\cite{KA}. Details of the model 
parameters and the reduced units used for length, time and temperature are the 
same as in \cite{KDS}. We have done simulations for $7$ different temperatures in
the range $T\in [{0.900,0.450}]$. For the 3dIPL model the system sizes studies are in 
the range $N\in [150, 10000]$ at number density $\rho = 1.20$ for $9$ different 
temperatures in the range $T\in[0.450,1.000]$. For the 3dR10 model system 
sizes studied are in range $N \in [64, 4000]$ at number density $\rho = 0.81$ 
for $9$ temperatures in the range $T\in[0.50,1.000]$. All the simulations are 
done in the canonical ensemble using a modified leap-frog integration
scheme with the Berendsen thermostat. We have also performed simulation with 
another constant temperature simulation algorithm due to Brown and Clark 
\cite{brownClark}. The results do not depend on the exact algorithm used for 
integrating the equations of motion. The equilibration runs for all these 
temperatures are close to $100~ \tau$ (where $\tau$ is the $\alpha$-relaxation 
time estimated from the decay of the two-point density correlation function) 
and we have averaged the data over $32$ independent runs of length $100 ~\tau$.
For the results discussed in Sec. IV, we performed Brownian Dynamics simulations 
using the predictor-corrector algorithm given in \cite{bdref}.

In \cite{KDS}, the $\alpha$ relaxation time was identified by the
decay of a two-point density correlation function, as well as by
considering the location of the peak of $\chi_4(t)$. In order to
define a $\beta$ time scale, we first consider an analogous procedure
of locating a short-time feature in $\chi_4(t)$.  Indeed, as observed
in \cite{Haxton_Liu}, there exists a maximum in $\chi_4(t)$ at short
times, whose location and height depend on the system size, as shown
in Fig. \ref{x4BetaPeak}. However, we find that the identified time
scale increases linearly with the length $L$ of the system. 
This behaviour is shown in the inset of Fig.\ref{x4BetaPeak}. This
plot also shows that this time scale continues to increase with increasing
$L$ without showing any sign of saturation. These observations suggest
a phononic origin of this time scale.
Such a possibility is further supported by the fact that
this feature disappears when we use Monte Carlo simulations to calculate $\chi_4(t)$. 
We opt not to use this procedure for identifying $\tau_\beta$.

In \cite{Stein_Andersen_PRL}, a short time scale $\tau_\beta$ was
identified as the time at which an inflection occurs in a log-log plot of 
the mean squared displacement {\it vs.} time. We use the same
definition to evaluate $\tau_\beta$ as a function of system size and
temperature.  In Fig.~\ref{msd}, we show the mean squared displacement
(MSD) as a function of time in a log-log plot and its derivative. The
clear dip in the derivative and its variation with temperature
indicate that this estimation of $\tau_\beta$ can be done
unambiguously and without much uncertainty. We use this definition of 
$\tau_\beta$  in the present work. 

\section{Results}
\begin{figure}[!h]
\begin{center}
\hskip -0.5cm
\epsfig{file=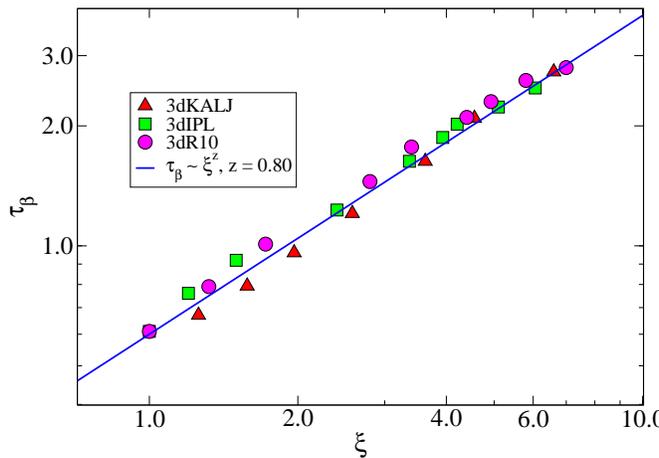,scale=0.410,angle=-90,clip=}
\caption{The relaxation time $\tau_\beta$, plotted as a function of
  the correlation length $\xi$, exhibits a power-law dependence, albeit
  over a small range of values. The line drawn through the data points represents
a power-law with exponent
  $z = 0.80$. }
\label{tauBetaXi}
\end{center}
\end{figure}
In Fig.~\ref{tauBetaSys}, we show the system-size dependence of
$\tau_\beta$ for different temperatures for the 3dKALJ model. For high temperatures the
dependence is weak and the asymptotic value is reached for small
system sizes, but this characteristic size becomes increasingly large
with decreasing temperature. Thus, $\tau_\beta$ shows a size dependence that is
opposite to that displayed by $\tau_\alpha$ \cite{KDS}, exhibiting
a system-size dependence that may normally be expected for a time scale
determined by an underlying length scale in the system. We
note that the range over which $\tau_\beta$ varies with system size
and temperature is modest, unlike the behaviour observed for $\tau_\alpha$.
Another important observation about the system-size dependence of $\tau_\beta$ is 
that the dependence of this time scale on the system size becomes negligibly small
if the system size is sufficiently large. This is very different from the behaviour of
the time scale associated with the short-time peak of $\chi_4(t)$ shown in
Fig.~\ref{x4BetaPeak}. As shown there, this time scale, which we attribute to
phonon-like excitations, continues to increase linearly with the length $L$ of
the system without showing any sign of saturation at large values of $L$.

We next scale the linear system size $L$ at each temperature by 
an empirically determined length
$\xi(T)$ such that the values of $\tau_\beta(N,T)/\tau_\beta(N\to \infty,T)$ for all
$N$ and $T$ ($\tau_\beta(N\to \infty,T)$ is the large-$N$ asymptotic
value of $\tau_\beta(N,T)$) collapse onto a master curve when plotted versus $L/\xi(T)$. 
The data collapse obtained this way, shown in Fig.~\ref{tauBetaSys}, is very good, 
This allows for the determination of a $\beta$-regime length scale $\xi(T)$.  

We then compare the length scale estimated in this way with the
heterogeneity length estimated in \cite{KDS} from quantities obtained at
$\tau_\alpha$. This comparison is shown in the inset of the 
bottom panel of Fig.~\ref{tauBetaSys}. 
In order to make the comparison, we scale the present estimate so that
the two length scales match at $T = 0.6$. Length scales obtained by
FSS are known only up to a multiplicative factor, and hence this
procedure does not introduce any additional arbitrariness. We find
that the temperature dependence of the two length scales agrees very
well. IMCT predicts that $\xi$ grows as a power law in time up to
$\tau_\beta$ and stays at the value $\tau_\beta$ up to
$\tau_\alpha$. Our results therefore confirm this expectation.
Independently of IMCT predictions, the agreement is remarkable in
pointing to an intimate connection between dynamics at short times
($\beta$ regime) and long times ($\alpha$ regime). This result
suggests that the heterogeneity present in the dynamics in the
$\alpha$ regime has been built up already in the $\beta$ regime, and
therefore in principle, essential information about the $\alpha$
relaxation can be obtained by studying short-time dynamics.  Our
result therefore lends support to many investigations and ideas that
aim to relate dynamical behaviour at short times to long-time structural
relaxation\cite{hallwolynes,buchenau,dyre,harrowellnphys,wyart,starr,leporini,douglas}. 

To see whether the results obtained for the 3dKALJ model are generic, 
we have done similar analysis for the 3dIPL and 3dR10 models.
In Fig. \ref{3dIPLresult}, we have shown, in the top left panel, the system-size
dependence of the peak value of the four-point susceptibility for different 
temperatures for the 3dIPL model. The top right panel shows the results of
FSS performed for the same
data to obtain the dynamic heterogeneity length scale. In the bottom left panel 
we show the system-size dependence of the short time scale
$\tau_{\beta}$ for different temperatures  and in the bottom right panel 
the corresponding FSS of $\tau_{\beta}$. The scaling
collapse observed in this case is also very good. We find that the temperature
dependence of the length scale obtained from the FSS of $\tau_{\beta}$ matches
quite well with the dynamic heterogeneity length scale, as shown in the inset of
the bottom right panel of Fig.\ref{3dIPLresult}. Similar analysis done
for the 3dR10 model  confirms that the same observations hold for this 
model too, as shown in Fig.\ref{3dR10result}.  

Finally we consider the dependence of $\tau_\beta$ on the extracted
length scale $\xi$.  In Fig.~\ref{tauBetaXi}, we show the relaxation
time as a function of the extracted correlation length in a log-log plot.
In contrast to $\tau_\alpha$ which exhibits deviations from a power-law
dependence on the corresponding length scale \cite{KDS}, we find
that a power law relation $\tau \sim \xi^{z}$, holds for $\tau_\beta$.
Although such a dependence is in qualitative agreement with IMCT, we
find $z \simeq 0.80$ which is at variance with the IMCT
prediction\cite{inhmct}. 



\section{Dependence on Microscopic Dynamics}

\begin{figure}[!h]
\begin{center}
\epsfig{file=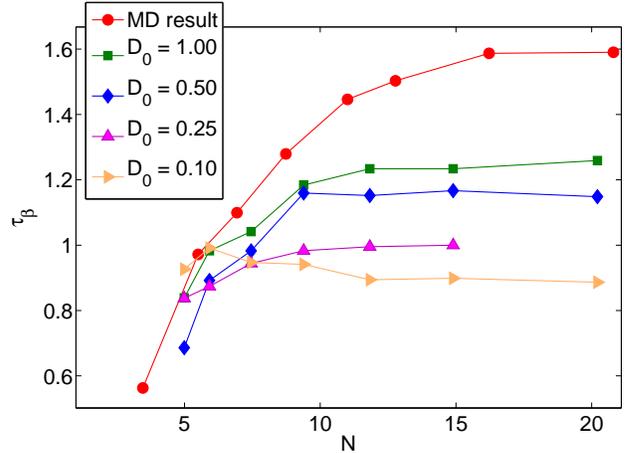,scale=0.40,clip=}
\caption{System-size dependence of the short-time $\beta$-relaxation time scale 
$\tau_{\beta}$ at $T = 0.470$ for the 3dKALJ model glass former from Brownian
dynamics simulations with different values  of the friction coefficient. $D_0$ 
indicated in the legend is
the microscopic diffusivity of the particles and is inversely proportional 
to the damping coefficient. Notice that for small damping (large $D_0$) the
system-size dependence is very similar to that obtained
in molecular dynamics simulations. The system-size dependence becomes weaker
with increasing damping. The time scale for Brownian dynamics is scaled
to match the long-time mean squared displacement for both molecular
dynamics and Brownian dynamics.}
\label{tauBetaCompare}
\end{center}
\end{figure}

In this section we present results concerning the dependence of the
time scale $\tau_{\beta}$ on the details of the microscopic dynamics. 
To address this issue systematically, we have performed Brownian dynamics (BD) simulations using 
a predictor-corrector scheme \cite{bdref}. This simulation scheme allows one to 
change systematically the friction coefficient to go from the very low friction limit  
(close to a molecular dynamics simulation) to overdamped dynamics with large
friction (close to a Monte Carlo simulation). 
We have changed the friction in our simulations over one order of magnitude
and studied its effect on the $\beta$-relaxation time scale. We find that although a 
$\beta$ time scale can be identified unambiguously for all the values of the friction 
coefficients studied, the system-size dependence becomes weaker with increasing 
friction and almost completely goes away for the largest value of the friction coefficient
considered here. The system-size dependence of $\tau_{\beta}$ is shown in 
Fig.\ref{tauBetaCompare} for the 3dKALJ model at temperature $T = 0.470$ for 
system sizes in the range $N\in [150, 10000]$. However, it is important to note 
that the time scale itself remains well-defined with increasing friction. 
This is different from the behaviour of 
the time scale obtained from the short-time peak in $\chi_4(t)$ \cite{Haxton_Liu}: 
as discussed above, this time scale can not be defined for Monte Carlo dynamics 
because the short-time peak in $\chi_4(t)$ is not 
present for this dynamics. Our results are consistent with previous work comparing 
the results for different microscopic dynamics \cite{gleim}, 
which found that the behaviour in the early $\beta$ regime is affected by the microscopic dynamics. 

We do not have a full understanding of the observed influence of the microscopic dynamics on the system-size dependence of $\tau_\beta$. There are reasons to
believe that the observed behaviour is connected to the effects of the properties of the inherent structures (local minima of the
potential energy), whose basins of attraction are visited by the system during its time evolution, on its dynamics. The dynamics in the short-time $\beta$-relaxation 
regime at low temperatures is expected to be strongly influenced by the properties of the inherent structures because the system should remain confined
in the basin of a single inherent structure (or in a single metabasin~\cite{heuer}, depending on the temperature) during its evolution over
relatively short times. A recent study~\cite{pranab} has shown that the dynamics of the 3dKALJ model in the $\beta$ relaxation regime (up to time
scales that are relatively short, but longer than the $\tau_\beta$ considered in our work), observed in molecular dynamics simulations at
temperatures near and below the glass transition temperature $T_c$ of mode coupling theory ($T_c \simeq 0.435$ for the 3dKALJ model),
can be understood from the low-energy properties of the relevant inherent structures. These properties include the eigenvalues and
eigenvectors of the Hessian matrix evaluated at the potential energy minimum (these define the ``normal modes'' of small-amplitude oscillations
near the bottom of the basin of an inherent structure) and the third and fourth derivatives of the potential energy at the minimum (these coefficients
determine the effects of anharmonicity on the normal modes). Thus, the system-size dependence of $\tau_\beta$ obtained in our Newtonian
molecular dynamics simulations and the length scale we have extracted from this dependence should be closely related
to these properties of the inherent structures. Our observation that the length scale extracted from the system-size dependence of $\tau_\beta$ obtained
from molecular dynamics simulations is essentially the same as the length scale of dynamic heterogeneity at the $\alpha$-relaxation time scale then
suggests that the spatial structure of dynamic heterogeneity at time scales of the order of the $\alpha$-relaxation time is closely related to the aforementioned properties of the inherent structures. This suggestion receives strong support from several experimental~\cite{expts} and 
numerical~\cite{harrowellnphys,numerics1,numerics2} studies that have shown that the spatial structure of dynamic heterogeneity at the $\alpha$-relaxation time scale is closely
related to the structure of the eigenvectors associated with some of the low-lying eigenvalues of the Hessian matrix evaluated at the appropriate
inherent structures. All these observations suggest that the short-time ($\beta$-relaxation) and long-time ($\alpha$-relaxation)
dynamics observed in molecular dynamics simulations are closely related to each other through the low-energy properties of the relevant inherent
structures. In Brownian dynamics simulations with a large friction coefficient, the influence of the properties of the inherent structures on
the short-time dynamics is masked by the strong damping of the normal modes. This may reduce or completely eliminate some of the effects of the 
structural properties of the 
inherent structures on the dynamics. The system-size dependence of $\tau_\beta$ which, as argued above, is closely related to the low-energy
properties of the inherent structures, may be one of the effects that disappear in the presence of strong friction. These arguments provide a
rationalization of the observed effects of strong friction on the system-size dependence of $\tau_\beta$ obtained from Brownian dynamics
simulations and suggest that the low-energy properties of the inherent structures visited by the system during its time evolution play an
important role in both $\beta$ and $\alpha$ relaxation processes.

\section{Discussion} 
 
In this work, we have studied the system-size dependence of the $\beta$ relaxation time 
for three model liquids in three dimensions. In each case, we find that the time scale 
initially increases with increasing system size and saturates for large values of the system size, exhibiting behaviour conforming to usual expectations for the size dependence of a quantity 
that depends on a length scale. This is unlike the system-size dependence of the 
$\alpha$ relaxation time that is found to decrease with increasing system size \cite{KDS}. On the other hand, the length scale extracted from FSS of this $\beta$ relaxation time scale matches very well with the length scale extracted from the FSS of the four-point
susceptibility $\chi_4^P(T)$ measured at the $\alpha$-relaxation time scale. This is in 
agreement with expectations based on inhomogeneous mode coupling theory, as also our 
observation that the asymptotic value of the $\beta$ time scale, obtained for large systems, 
exhibits a power-law dependence on the length scale. The value of exponent of this dependence, however, 
does not match IMCT predictions ($\frac{4}{2 a} \simeq 5.4$) \cite{inhmct}. Our results are also consistent with  ideas that relate dynamics at short times to long-time structural
relaxation\cite{hallwolynes,buchenau,dyre,harrowellnphys,wyart,starr,leporini,douglas}. 
We have also presented intriguing results for the dependence of the short-time dynamics
on the microscopic dynamics used in the simulation and suggested a possible explanation
of the observed behaviour in terms of the role played by certain low-energy 
properties of the relevant inherent structures (i.e. inherent structures whose basins are 
visited by the system during its time evolution) in the dynamics of the system.

The results reported here have a strong connection with those in \cite{CKPS2012}, 
in which oscillatory shear simulations were performed to study the effects of
short-time $\beta$-relaxation in a supercooled two-dimensional glass-forming liquid 
at both high and low temperatures. The loss modulus measured in this study was 
shown to be related to $\beta$-relaxation and the effect of this relaxation was 
found to decrease sharply by the introduction of a small 
amount of pinning disorder in the system. The cooperative displacements of 
individual particles over the $\beta$-relaxation time scale were found to 
diminish very rapidly with the introduction of pinning disorder. Based on these 
observations, it was concluded  that $\beta$-relaxation is also cooperative in nature, 
similar to $\alpha$-relaxation. Our study clearly shows that $\beta$-relaxation is 
indeed cooperative and the cooperativity is likely to have the same origin as 
that in $\alpha$-relaxation.

\end{document}